\documentclass[prb,twocolumn,aps,showpacs]{revtex4}
\usepackage{graphicx}% complex graphics
\usepackage{bm}
\usepackage{amsmath,amssymb}
\usepackage{subfigure}
\usepackage{float}
\usepackage{latexsym}
\usepackage{epstopdf}

\begin{document}
\newcommand{\s}{\scriptscriptstyle}
\newcommand{\uu}{\uparrow \uparrow}
\newcommand{\ud}{\uparrow \downarrow}
\newcommand{\du}{\downarrow \uparrow}
\newcommand{\dd}{\downarrow \downarrow}
\newcommand{\ket}[1] { \left|{#1}\right> }
\newcommand{\bra}[1] { \left<{#1}\right| }

\title{Organic magnetoresistance near saturation: mesoscopic effects
in small devices}
\author{R. C. Roundy,  Z. V. Vardeny, and M. E. Raikh}
\affiliation{Department of Physics and Astronomy, University of Utah, Salt Lake City, UT 84112}

\begin{abstract}
In organic light emitting diodes with small area
the  current may be dominated by a {\em finite} number, $N$ of
sites in  which the electron-hole recombination occurs.
As a result, averaging
over the hyperfine magnetic fields, ${\bm b}_{\s h}$,
that are generated in these sites  by the environment nuclei
is incomplete. This creates
a random (mesoscopic) current component, $\delta I({\bm B})$,
at field ${\bm B}$ having  relative  magnitude $\sim N^{-1/2}$.
To quantify the statistical properties of  $\delta I({\bm B})$
we calculate the correlator
$K({\bm B}, {\s \Delta}{\bm B})=\langle\delta I({\bm B}-\frac{{\s \Delta}{\bm B}}{2})
\delta I({\bm B}+\frac{{\s \Delta}{\bm B}}{2})\rangle$
for parallel, ${\s \Delta}{\bm B} \|  {\bm B}$, and perpendicular, ${\s \Delta}{\bm B}\perp    {\bm B}$ orientations of ${\s \Delta}{\bm B}$.
We demonstrate that mesoscopic fluctuations develop at fields $|{\bm B}|\gg |{\bm b}_{\s h}|$, where the average magnetoresistance is near saturation. These fluctuations originate from the {\em slow}
beating between the singlet,$S$ and triplet, $T_{\s 0}$ states of the recombining $e$-$h$ spin pair-partners.
We identify the most relevant processes responsible for the current fluctuations
as due to anomalously slow
beatings that develop  in sparse $e$-$h$  polaron pairs at sites
for which  the ${\bm b}_{\s h}$ projections
on the external field direction almost coincide.

\end{abstract}
\maketitle

%%%%%%%%%%%%%%%%%%%%%%%%%%%%%%%
\section{Introduction}
%%%%%%%%%%%%%%%%%%%%%%%%%%%%%%%%

In the field of `Dynamic Spin Chemistry', a mechanism by which
the recombination rate
of radical pairs is sensitive to
a weak magnetic field, ${\bm B}$, was established
more than four decades ago; see, e.g., the reviews in Ref. \onlinecite{reviews}.
This mechanism relies on the hyperfine interaction of the spin-$1/2$
pair partners with their respective nuclear spin environments,
where the hyperfine field, ${\bm b}_{\s h}$ generated by the nuclei
is responsible for the radical spins dynamics in zero field.
In this process if at time $t=0$ the radical pair spin state
is, e.g., in a singlet configuration,  $S$,
then at finite $t$
it will acquire a triplet ($T$) component with  probability,
$P_{\s ST}(t)$.
If recombination is allowed only from $S$,
then $P_{\s ST}(t)$ dynamic evolution
affects the recombination rate.
Clearly, $P_{\s ST}(t)$ depends on ${\bm B}$ and
this sets a small scale, $|{\bm B}|\sim |{\bm b}_{\s h}|$
that may influence the radical pair recombination rate.

\begin{figure}
\includegraphics[width=77mm, angle=0,clip]{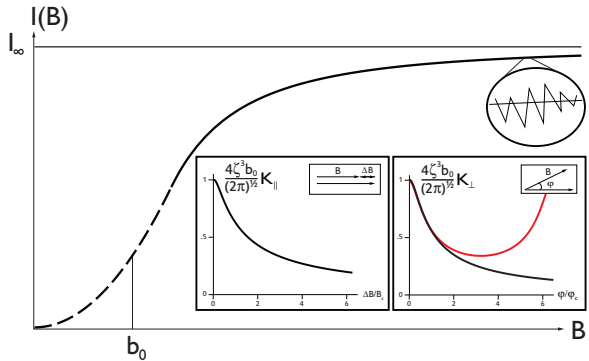}
\vspace{-0.4cm}
\caption{The dependence $I(B)$ of the device current on the applied magnetic field is
shown schematically in
the strong-field limit $B\gg b_{\s 0}$. Enlargement illustrates
mesoscopic fluctuations emerging in a small sample. Two insets are the correlators of
the mesoscopic fluctuations for ${\s \Delta}{\bm B} \parallel {\bm B}$ and
${\s \Delta}{\bm B} \perp {\bm B}$  plotted from Eqs. (\ref{PhiParallel}) and (\ref{PhiPerp}), respectively . }
\label{figOMAR}
\end{figure}

An important advance in the quantitative description
of $P_{\s ST}(t)$
 was made by Schulten and Wolynes\cite{Schulten}.
They  noticed that, due to the large
number of nuclei surrounding each radical pair,
and  slow dynamics of the hyperfine field,
the ${\bm b}_{\s h}$ random distribution may be modeled
by a  Gaussian.  Under these conditions
the multiplicity of the nuclear
spin configurations
may be  characterized
by a single number -- namely the width of this
distribution, $b_{\s 0}$.

\begin{figure}
\includegraphics[width=77mm, height=50mm, angle=0,clip]{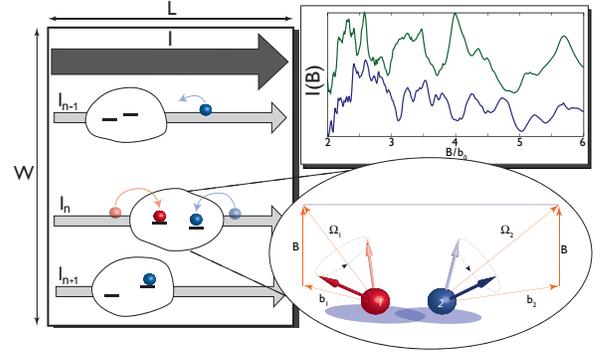}
\vspace{-0.4cm}
\caption{In strongly inhomogeneous device with $W\gg L$ the current passage
is dominated by the most conductive channels, $I=\sum_{n}I_n$.
Each current component is limited by the most resistive junction, illustrated schematically. The
current through this junction is sensitive to the spin dynamics of the constituting PP.
``Slow'' pairs, shown in the enlargement, are those in which the $z$-projections
of their hyperfine fields coincide accidentally. The inset shows
 $\sum_{n=1}^{N}I_n$ calculated for two realizations of $N=10^4$ random hyperfine fields with rms $b_0=10^{2}\zeta^{-1}$.  }

\label{figsample}
\end{figure}

The dependence of $P_{\s ST}(t)$ on ${\bm B}$
is at the core of organic magnetoresistance (OMAR),
which has recently attracted a lot of attention\cite{Markus0,Markus1,Markus2,Prigodin,Gillin,Valy0,Valy1,Valy2,Valy3,Blum,Bobbert1,Bobbert2,Wagemans2,Wagemans,Wagemans1,Flatte1}.
This is because the
current, $I$ in a biased organic diode
involves recombination of the injected $e$-$h$
polarons (forming polaron pairs, PP); whereas the
processes of populating and depopulating of traps
are not sensitive to spin dynamics.

The theory of OMAR is conceptually harder than that of
spin-magneto-chemistry\cite{reviews} for two reasons.
Firstly, in OMAR  the complex dynamics
of all four  PP spin states, $S$, $T_{\s 0}$, $T_{\s +}$, and $T_{\s -}$
needs be incorporated into the calculation of the dc current
that is influenced by the PP.
Secondly, each PP is sensitive to the other PPs if they
belong to the same current path.
Finally, averaging over the nuclear environment should
be carried out only at the last step.
To bypass these complications several
simplifying assumptions concerning both the spin dynamics
and current passage scenario were adopted
in previous theoretical calculations
of the $I(B)$ response\cite{Prigodin,Blum,Bobbert1,Bobbert2,Wagemans2,Wagemans,Wagemans1,Flatte1}.

In contrast, in the present paper we do not focus  on the entire $I(B)$ response,
but rather on the strong $B$ domain, where the OMAR response
is close to saturation, see Fig. \ref{figOMAR}.
Our motivation is twofold. Firstly, theory allows
a dramatic simplification in this $B$-domain, since the
spin dynamics that is relevant to OMAR involves only the PP  $S$ and $T_{\s 0}$ states.
However even in this $B$-domain the OMAR underlying
physics is not trivial if the hyperfine field is sufficiently strong;
namely when $b_{\s 0}\tau \gg 1$, which corresponds to the regime of ``slow'' hopping\cite{Wagemans,Flatte1}.
Here $\tau$ is a characteristic recombination time, and $b_{\s 0}$ is measured
in frequency units. Experimentally\cite{Bobbert1},
in organic semiconductors $b_{\s 0}$ is $\sim 1\,\text{mT}$, whereas
$\tau \sim 1-10\,\mu\text{s}$, so that this parameter
is $\sim 10^3$.  We show that at large $b_{\s 0}\tau$
the spin dynamics is not ``frozen'' as $B$ exceeds $b_{\s 0}$, but persists in a parametrically broad interval,
$b_{\s 0}^2\tau \gg B\gg b_{\s 0}$.
Our second and central motivation for considering strong fields is that we
predict the occurrence of {\em mesoscopic} properties in this $B$-domain that
would form in small devices that are based on strongly disordered organic active
layers. Specifically we predict
{\em reproducible} random fluctuations in the $I(B)$
response upon sweeping $B$ (see Fig. \ref{figOMAR}),
which reflect the ``individuality''\cite{Matveev,review,mesoscopic} of the nuclear
environments associated with the relevant recombination centers
in the organic.

More quantitatively, if the number, $N$ of current paths
that contribute to $I(B)$ is finite, then the statistical averaging
over ${\bm b}_{\s h}$ is incomplete. The relative fluctuation
$\frac{\delta I(B)}{\langle I \rangle} \sim N^{-1/2}$ while small,
can  be still experimentally obtained because of the
high accuracy with which current
can be measured.
In the field of `dynamic spin chemistry', mesoscopic fluctuations cannot
occur since
the number of radical pairs that contribute to the observable characteristics
is huge.

Obviously, the necessary condition to observe  mesoscopic
fluctuations in the $I(B)$ response of organic devices is
slow nuclear spin dynamics,
which should allow one to obtain $I(B)$
before the nuclear configuration changes.
This is realistic, since the characteristic time
for current passage is a PP recombination time, which
for organic devices is $50 \mu\text{s}$, see Ref. \onlinecite{Dane}.
It is generally accepted\cite{Schulten,Loss} that the time for
the change of the nuclear-spin configuration is orders of magnitude
longer, although no accurate measurements of proton spin-spin relaxation
time for organic devices have yet been reported in the literature.

\section{PP dynamics in strong fields}

\subsection{Isolated PP}
We start with a detailed account of the PP spin dynamics and recombination in the strong $B$-domain,
which we then use to calculate
mesoscopic contribution to $I(B)$ near  saturation.
%\noindent{$S-T_{\s 0}$ {\em beating in the strong-field limit.}}
For an isolated PP the spin Hamiltonian
$\widehat{H}={\bm \Omega}_1\cdot \widehat{{\bm S}}_1+{\bm \Omega}_2\cdot \widehat{{\bm S}}_2$ describes the
precession of the PP spins ${\bm S}_1$, ${\bm S}_2$ in the fields ${\bm \Omega}_1={\bm B}+{\bm b}_1$ and
${\bm \Omega}_2={\bm B}+{\bm b}_2$, respectively. If at $t=0$ the PP is in the singlet state,
then the probability, $P_{\s SS}(t)$ to find it in the singlet state at finite $t$ oscillates with time.
 $P_{\s SS}(t)$ oscillations contain two frequencies: $\Delta=|{\bm \Omega}_1|-|{\bm \Omega}_2|$ and $\Sigma=|{\bm \Omega}_1|+|{\bm \Omega}_2|$.
The advantage
in considering the strong-field limit is that since $|{\bm \Omega}_1|\approx |{\bm \Omega}_2|$,
the frequencies $\Delta$ and $\Sigma$  are very different from each other, so that the spin dynamics
decouples into distinct `slow' and `fast' modes. Moreover, the slow mode
involves predominantly $S$ and $T_{\s 0}$ states, while the admixture of $T_{\s +}$ and
$T_{\s -}$ states to this mode is relatively weak (of the order of $b_{\s 0}^2/B^2$).
The fast mode $\Sigma$ has frequency $\approx 2B$ and describes the oscillations between $S$ and $T_{\s +}$, $T_{\s -}$.
But the admixture of $S$ to this mode is also suppressed as $b_{\s 0}^2/B^2$ in the strong-field limit.
We thus conclude that, with accuracy $b_{\s 0}^2/B^2$,  $P_{\s SS}(t)$ dynamics simplifies in the
strong-field limit to
$P_{\s SS}(t)=\cos^2\Delta t$; namely the `beating' between $S$ and $T_{\s 0}$ states.
Similarly, if in the strong-field limit
the PP is initially in the $T_{\s 0}$ state then the probability to find it in the $S$ state at time $t$ is
$\sin^2\Delta t$.

\subsection{ Recombination in the presence of $S-T_{\s 0}$  beating}

We now assume  that the PP is still isolated from the `leads',
but can recombine from $S$ to the ground state, $G$.
A crucial question for OMAR is: what are the waiting times
$\langle t \rangle_{\s S}$, $\langle t \rangle_{\s T_{\s 0}}$
for the recombination, if the system is initially in $S$ and $T_{\s 0}$,
respectively. The simplified spin dynamics in the strong-field limit
allows us to address this question analytically.

Upon restricting the basis  to $S$, $T_{\s 0}$ and  the ground state,
we have $9$ relevant elements of the density matrix for
solving the Liouville-Lindblad equations of motion:
$\dot{\rho} = -i [\hat{H},\rho] + \hat{L}(\rho)$, where the
operator $\hat{L}(\rho)$ describes the recombination. To find, e.g.
$\langle t \rangle_{\s S}$ the system should be solved with the
initial conditions $\rho(0) =\ket{S}\bra{S}$. Subsequently
$\langle t \rangle_{\s S}$ is found from the formula
\begin{equation}
\label{formula}
\langle t \rangle_{\s S}=\int_0^{\infty}\!\!dt~t~ \frac{\partial{\rho_{\s GG}}}{{\partial t}}=\int_0^{\infty}dt\, \bigl(\rho_{\s SS}(t)+\rho_{\s T_{\s 0}T_{\s 0}}(t)\bigr).
\end{equation}
Similarly  $\langle t \rangle_{\s T_{\s 0}}$ is obtained
from Eq. (\ref{formula}) upon solving the equations
of motion with initial conditions $\rho(0) =\ket{T_{\s 0}}\bra{T_{\s 0}}$.
These calculations yield

\begin{equation}
\label{times}
\langle t \rangle_{\s S}=\tau, ~~~~~
\langle t \rangle_{\s T_{\s 0}}=\tau + \frac{1}{2 \tau \Delta^2}.
\end{equation}
For a typical PP we have $\Delta= |{\bm \Omega}_1|-|{\bm \Omega}_2|\sim b_{\s 0}$.  Eq. (\ref{times})
suggests that
$\langle t \rangle_{\s S}\approx\langle t \rangle_{\s T_{\s 0}}\approx \tau$.
This  is a natural result since recombination is preceded by many beatings
between $S$ and $T_{\s 0}$ states; therefore the recombination time does not
depend on the  initial PP state. The most
striking consequence of Eq. (\ref{times}) is that for {\em sparse}
PP for which $\Delta$ is accidentally smaller than $\tau^{-1}$
we have $\langle t \rangle_{\s S}\ll \langle t \rangle_{\s T_{\s 0}}\approx \frac{1}{2\Delta^2\tau}$.
This suggests that the smaller is $\Delta$,
the longer the pair stays ``trapped'' in $T_{\s 0}$. Note that
in the course of beating {\em without possibility of recombination}, such PP
would cross from $T_{\s 0}$ to $S$ after much shorter time
$\Delta^{-1} \ll \langle t \rangle_{\s T_{\s 0}}$.
We can trace the origin of the ``trapping'' described by Eq. (\ref{times}) to the complex eigenmodes of
the system that consists of singlet and triplet components being mixed by the hyperfine field. This system may be described by the $2\times 2$ non-hermitian matrix:
\begin{equation}
\label{scary}
\bordermatrix{  & \ket{S} & \ket{T} \cr
 & -i/\tau & \Delta \cr
 & \Delta & 0 \cr
},
\end{equation}
where the nondiagonal elements describe the mixing, while $-i/\tau$
describes recombination from $S$ to $G$. The eigenvalues of this matrix are
\begin{equation}
\label{lambda}
\lambda_{\s 1,2}=-\frac{i}{2\tau}\pm \sqrt{\Delta^2-\frac{1}{4\tau^2}}.
\end{equation}
In the limit $\Delta \ll \tau^{-1}$ we have $\lambda_{\s 1}\approx -\frac{i}{\tau}$,
while $\lambda_{\s 2}\approx i\tau\Delta^2$. We see that $\lambda_{\s 2}$ is anomalously small,
and the result Eq. (\ref{times}) for $\Delta \ll \tau^{-1}$ can be interpreted as
$\langle t \rangle_{\s T_{\s 0}}\sim \frac{1}{\lambda_{\s 2}}$.
We note in passing that the emergence of slow mode, $\lambda_{\s 2}$ in a compound system
with anomalously close levels  was previously found in
Refs. \onlinecite{subradiance1,Gefen,subradiance2} in connection with
resonant tunneling through pairs of localized states.

\begin{figure}
\includegraphics[width=77mm, angle=0,clip]{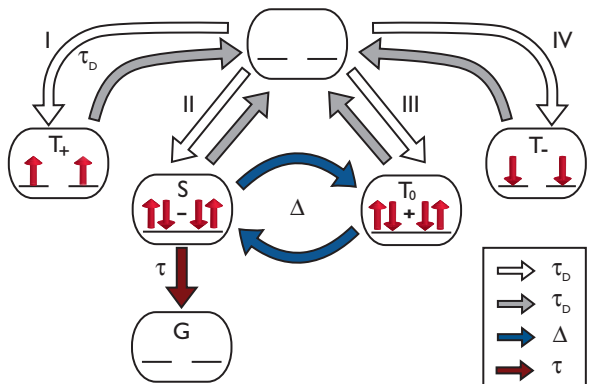}
\vspace{-0.4cm}
\caption{Schematic representation of the PP population dynamics in the strong-field domain.
For variants of cycle I (IV)  the pair is assembled and, subsequently, disassembled in $T_{\s +}$ ($T_{\s -}$) state. For variants II (III) the pair is assembled in $S$ ($T_{\s 0}$) state in which it undergoes slow dynamics prior to disassembly. In the course of the slow dynamics the pair can
recombine; recombination is possible {\em only} from $S$.
Since the transport is {\em unidirectional},
current is passed through a junction upon completion of each  cycle variant. }
\label{figcycle}
\end{figure}

Eqs. (\ref{scary}) and (\ref{lambda}) provide a semi-qualitative derivation of
our main result Eq. (\ref{times}).  A detailed derivation which justifies the above procedure
is presented in Ref. \onlinecite{WE} for the case of arbitrary external field. In particular, this paper deals with a delicate fact that the eigenvectors of non-hermitian  matrix Eq. (\ref{scary}) are not orthogonal to each other.

In the remainder of the paper we demonstrate that it is the sparse pairs with $\Delta \lesssim \tau^{-1}$ that are responsible for
 the mesoscopic part of the  $I(B)$ response in the strong-field limit.
%as well as  for the mesoscopic
%fluctuations in  $I(B)$ response.

\section{ Transport model}

We assume that the organic active layer in the device
is strongly inhomogeneous and its width, $W$ is much larger than the distance,
$L$ between the electrodes,  see Fig. \ref{figsample}.
Under these conditions the minimal description of transport is to model the
 sample as  $N\gg 1$ parallel conducting channels. Moreover due to the film
inhomogeneity, the current through each channel is limited
by a single, {\em most resistive} junction. The stronger is the inhomogeneity,
 the more realistic is the proposed model, see e.g. the review Ref. \onlinecite{review}.
The net current through the sample is the sum, $I=\sum_{n=1}^N I_n$, of the currents in each channel.
Each junction, $n$, can be viewed as a pair of sites coupled to the nuclei environment.
In the course of the current  $I_n$ through the junction,
the pair of sites first gets occupied, is then emptied, and so on.
In other words, the current passage  can be viewed as a sequence
of {\em cycles}, see Fig. \ref{figcycle}. Each cycle consists of two steps,
namely {\em assembly} of a pair on neighboring sites and disappearance of the
pair through either dissociation or recombination, see Fig. \ref{figcycle}.
At this point we emphasize that it is the recombination
stages of the cycles II and III (defined in Fig. \ref{figcycle})
that are described by Eq. (\ref{times}), and are thus sensitive to $B$.

The simplified  transport model described in Fig. \ref{figsample} encodes the same picture of transport put forward in Ref. \onlinecite{Bobbert1}.
It contains recombination and bypassing, if recombination takes too long.
From Fig. \ref{figcycle} we may write the average duration $\langle t_{\s n} \rangle$
of the cycle as follows
\begin{equation}
\label{cycle}
\langle t_{\s n} \rangle = 2 \times \frac{1}{4}(2 \tau_{\s D}) + \frac{1}{4}(\langle t \rangle_{\s S} +\tau_{\s D})+\frac{1}{4}(\langle t \rangle_{\s T_{\s 0}} +\tau_{\s D}).
\end{equation}
The first term in Eq. (\ref{cycle}) originates from
the variants I and IV of the current cycle when the pair is assembled, and subsequently disassembled in $T_{\s +}$ and $T_{\s -}$ states, respectively. Then the cycle lasts for time $2\tau_{\s D}$.
The last two terms in Eq. (\ref{cycle}) describe the current
cycle variants II and III, in which the pair is assembled in either $S$ or  $T_{\s 0}$.  Eq. (\ref{cycle}) takes into account that realization of each current cycle has equal probability of $\frac{1}{4}$. For simplicity we assume that  processes involving leads, namely, assembly and dissociation, take equal time, $\tau_{\s D}$.
The current through the junction, which is the inverse cycle duration, can be then cast in the form
\begin{equation}
\label{current}
I_{\s n} = \frac{1}{\langle t_{\s n} \rangle}=8\zeta^2\tau -\delta I_{\s n}({\bm B}), ~~
\delta I_{\s n}({\bm B})=\frac{8\zeta^4\tau}{\Delta_{\s n}{({\bm B})}^2+\zeta^2},
\end{equation}
where we used Eq. (\ref{times}) and introduced the characteristic frequency
\begin{equation}
\label{zeta}
\zeta=\frac{1}{2\bigl[\tau(3\tau_{\s D}+\tau)\bigr]^{1/2}}.
\end{equation}
We emphasize that the correction, $\delta I_{\s n}({\bm B})$, in Eq. (\ref{current})
originates from $S-T_{\s 0}$ beating. For a {\em typical} nuclear environment we have $\Delta_{\s n} \sim b_{\s 0}$, so that the relative magnitude of this correction is $\sim \frac{\zeta^2}{b_{\s 0}^2} \ll 1$.
 However,
{\em on average}, this term is  much bigger, since it is dominated by sparse configurations with anomalously
small $\Delta_{\s n} \sim \zeta$. This is because, while the portion of
these sparse configurations is small, $\sim \zeta/b_{\s 0}$, the  $\delta I_{\s n}$ value
for these configurations exceeds the typical $\delta I_{\s n}$ by a large
factor $\frac{b_{\s 0}}{\zeta}$.  It can be demonstrated through a careful analysis\cite{WE}
that the correction $\delta I_{\s n}({\bm B})$ is insensitive to ${\bm B}$ {\em on average}.
The number of ``slow" pairs decreases with $B$. It was established in Ref. \onlinecite{WE}
that the PP recombination time also decreases with $B$ in such a way that the two tendencies
compensate each other identically.
In spite of this, it is the  correction Eq. (\ref{current})  that gives rise to the mesoscopic fluctuations of current to which we
now turn.

\section{ Mesoscopic fluctuations}

If a {\em given}  pair
contributes to the correction $\delta I_{\s n}({\bm B})$ in  Eq. (\ref{current}),
then the $S-T_{\s 0}$ splitting, $\Delta_{\s n} ({\bm B})$  for this pair is
$\sim \zeta$. This suggests that, upon changing ${\bm B}$ by a small
${\s \Delta} {\bm B}$, the condition  $\Delta_{\s n} ({\bm B})\sim \zeta$ for {\em this} pair is violated, while it becomes satisfied for {\em different} pairs. Such
``switching'' of pairs contributing to the correction, $\delta I(\bm {B})$
gives rise to the mesoscopic fluctuations of the current, which we may
quantify by the correlator
\begin{equation}
\label{K}
K({\bm B}, {\s \Delta}{\bm B})=\left< \delta I({\bm B}\!-\frac{{\s \Delta}{\bm B}}{2})
\delta I({\bm B}+\frac{{\s \Delta}{\bm B}}{2})\right>\!-\!\langle  \delta I({\bm B}) \rangle^2.
\end{equation}
 We  consider two cases.  In the first case, ${\s \Delta}{\bm B} \parallel {\bm B}$, the two magnetic fields are collinear.  In  the second case, ${\s \Delta}{\bm B} \perp {\bm B}$, the two magnetic fields have the same magnitude but are rotated through an angle $\varphi$ with respect to each other.

For calculating the correlator in Eq. (\ref{K}) we recast the factor $(\Delta^2+\zeta^2)^{-1}$ in $\delta I_{\s n}$
as a Fourier transform
\begin{equation}
\label{fourier}
\frac{1}{\Delta^2 + \zeta^2} = \frac{1}{2\zeta^2} \int_{-\infty}^\infty ds \exp\left( - |s| + i\frac{s\Delta}{\zeta} \right).
\end{equation}
By virtue of this transformation, the beating frequency $\Delta$, which depends on
the hyperfine fields, appears in the exponent of the integrand. Next we
take advantage of the fact that the beating
frequency, $\Delta$ in the strong-field limit  can be expanded as
\begin{equation}
\label{Delta}
\Delta = ({\bm b}_{\s 1} - {\bm b}_{\s 2}) \cdot {\bm n} +
 \frac{ {\bm b}_{\s 1}^2- {\bm b}_{\s 2}^2- ({\bm b}_{\s 1} \cdot {\bm n})^2
  +({\bm b}_{\s 2} \cdot {\bm n})^2}{2B},
\end{equation}
where ${\bm n}$ is the unit vector in the direction of ${\bm B}$.
Now, since $\Delta$ contains only linear and quadratic terms in ${\bm b}_i$,
the averaging of the exponential factor can be performed explicitly using
the properties
\begin{equation}
\label{auxiliary}
\left< e^{i \kappa b_{\s i}} \right> = e^{-\kappa^2 b_{\s 0}^2 /4}, \quad
\left< e^{ -\mu b_{\s i}^2 } \right> = \frac{1}{\sqrt{1 + \mu b_{\s 0}^2}}.
\end{equation}
For the parallel case, this averaging yields
\begin{widetext}
\begin{equation}
\label{Kparallel}
\left< \frac{1}{(\Delta({\bm B}+\frac{{\s \Delta}{\bm B}}{2})^2 + \zeta^2)
    (\Delta({\bm B}-\frac{{\s \Delta}{\bm B}}{2})^2 + \zeta^2)} \right>
    = \frac{1}{4\zeta^4} \int\limits_{-\infty}^\infty \!\!\!ds_{\s 1}
        \!\!\int\limits_{-\infty}^\infty \!\!\!ds_{\s 2}
          \frac{\exp\left(-(|s_{\s 1}|+|s_{\s 2}|) -\frac{b_{\s 0}^2}{2\zeta^2}\left(s_{\s 1} -s_{\s 2} \right)^2 \right)}{
    1 + \frac{b_{\s 0}^4}{4 \zeta^2}\left(\frac{s_{\s 1}}{B+\frac{{\s \Delta B}}{2}} - \frac{s_{\s 2}}{B-\frac{{\s \Delta B}}{2}} \right)^2}.
\end{equation}
\end{widetext}
The exponent $\exp\left[-\frac{b_{\s 0}^2}{2\zeta^2}\left(s_{\s 1} -s_{\s 2} \right)^2\right] $
follows from the first identity Eq. (\ref{auxiliary}), while the denominator emerges from
the second identity.

As a next step we perform the integration over the difference $s_{\s 1}-s_{\s 2}$. This integration can be performed explicitly using the fact that $\zeta \ll b_{\s 0} \ll B$.  Upon this integration, the average
Eq. (\ref{Kparallel}) can be presented in the form $\frac{\sqrt{2 \pi}}{4\zeta^3b_{\s 0}}\Phi_{\parallel}(\frac{{\s \Delta}B}{B_{\s c}})$, where the dimensionless
function $\Phi_{\parallel}$ is defined as
\begin{equation}
\label{PhiParallel}
\Phi_{\parallel}(z)=\int_0^{\infty}dx~ \frac{e^{-x}}{1+z^2x^2},
\end{equation}
and $B_{\s c}=4B^2\zeta/b_{\s 0}^2$.
The argument of the function $\Phi$ imposes a characteristic ``period" of mesoscopic fluctuations: $\delta B \sim \frac{B^2\zeta}{b_{\s 0}^2}$.

Eq. (\ref{PhiParallel}) suggests that  the period of the mesoscopic fluctuations  grows quadratically with  $B$. We tested this result by a numerical simulation.
For this simulation we chose $N=10^4$ random values of ${\bm b}$ with rms $b_0=10^{2}\zeta^{-1}$.
For each set of the local hyperfine fields  the sum  $\sum_{n=1}^{N}\delta I_n({\bm B})$, where $\delta I_n$ is given by Eq. (\ref{current}) was evaluated. The results of simulation are shown in
Fig. \ref{figsample}. Mesoscopic fluctuations and growth of their period with $B$ are apparent.

Our consideration applies for $\delta B \ll B$. This  condition
suggests that for measuring  the fluctuations
one must work in the domain $b_{\s 0} \ll B \ll b_{\s 0}^2/\zeta$.
The correlator Eq. (\ref{PhiParallel})
 is plotted in Fig. \ref{figOMAR}.
For small $\delta B \ll B_{\s c}$ it behaves as
$1-(\frac{\delta B}{B_{\s c}})^2$, and falls off slowly, as $\frac{\pi B_{\s c}}{2\delta B}$, for $\delta B \gg B_{\s c}$.

For the perpendicular case, we can simplify $\Delta$ as $\Delta \approx
{({\bm b}_{\s 1} - {\bm b}_{\s 2})\cdot {\bm n}}$. This is because the ${\bm B}$-dependence of $\Delta$ enters via the orientation,
${\bm n}$. Performing the same decoupling (Eq. (\ref{fourier})) as for the parallel case, instead of the
double integral in Eq. (\ref{Kparallel}) we get now
\begin{widetext}
\begin{equation}
\label{Kperp}
\left< \frac{1}{(\Delta({\bm B}+\frac{{\s \Delta}{\bm B}}{2})^2 + \zeta^2)
    (\Delta({\bm B}-\frac{{\s \Delta}{\bm B}}{2})^2 + \zeta^2)} \right>
    =
\int \!\!\frac{ds_{\s 1}}{2\zeta^2} e^{-|s_{\s 1}|} \!\! \int \!\!\frac{ds_{\s 2}}{2 \zeta^2} e^{-|s_{\s 2}|}\!
\exp\!\left\{\!\!-\frac{b_{\s 0}^2}{2 \zeta^2}
\!\left(s_{\s 1}^2\!+\!s_{\s 2}^2\!-\!2s_{\s 1}s_{\s 2}\cos\varphi\right)\right\}.
\end{equation}
\end{widetext}
We again see that by virtue of the relation
 $b_{\s 0} \gg \zeta$, the difference $(s_{\s 1} - s_{\s 2}) \sim \frac{\zeta}{b_{\s 0}}$ is small.  This allows us to integrate over $s_{\s 1} - s_{\s 2}$ and reduce Eq. (\ref{Kperp}) to $\frac{\sqrt{2\pi}}{4 \zeta^3 b_{\s 0}}\Phi_{\perp}(\frac{\varphi}{\varphi_{\s c}})$,
 where $\varphi_{\s c}=\frac{2 \sqrt{2} \zeta}{b_{\s 0}} \ll 1$ and the function $\Phi_{\perp}(z)$ is defined as
 \begin{equation}
 \label{PhiPerp}
 \Phi_{\perp}(z)=\int_0^{\infty}dx~ \exp\bigl[-x-z^2x^2\bigr].
 \end{equation}
The correlator  is plotted in Fig. \ref{figOMAR}. At $\varphi\gg \varphi_{\s c}$, it falls off as $\varphi_{\s c}/\varphi$.
In general, the correlator, Eq. (\ref{Kperp}) is a periodic
function of $\varphi$; had we not used the small-$\varphi$
expansion it would go through a minimum at $\varphi= \pi/2$ and
``revive'' at $\varphi=\pi$.

Our results related to mesoscopics can be summarized in the following
expression
\begin{equation}
\frac{\left< \delta I\left({\bm B}-\frac{\Delta {\bm B}}{2}\right)
\delta I\left({\bm B}+\frac{\Delta {\bm B}}{2}\right) \right>}{I(\infty)^2} =
\frac{\sqrt{2 \pi} \zeta}{4 N b_{\s 0}}
\left\{  \begin{matrix} \Phi_{\parallel} \left( \frac{\Delta B}{B_{\s c}} \right), \, {\s \Delta}{\bm B} \parallel {\bm B} \\
\Phi_{\perp} \left( \frac{\varphi}{\varphi_{\s c}} \right), \, {\s \Delta}{\bm B} \perp {\bm B} \end{matrix} \right.
\end{equation}

\section{Discussion}
\begin{itemize}

\item
By choosing a  simple  transport model for an organic semiconductor device,
and  adopting the  assumption\cite{Bobbert1,Flatte1} that
recombination proceeds exclusively from the singlet state,
we
were able to  demonstrate mesoscopic fluctuations in the OMAR response
in the domain $B\gg b_{\s 0}$, where the average current is saturated,
%capture
%analytically the OMAR response in the
%domain $B\gg b_{\s 0}$,
and predict their characteristic magnitude and period.
% of mesoscopic fluctuations.
Our theory is based on an
%important finding
observation that
in this $B$-domain there exists a strong separation between slow and fast
components of the PP spin-dynamics. As a result of this separation,
the $S$-$T_{\s 0}$ beating becomes decoupled, which, in turn, leads to a dramatic recombination
slow down which originates from PP  ``trapping''  in the $T_{\s 0}$ state.
Since this underlying physics is so general, any transport model in a small device
with few junctions  should exhibit mesoscopic fluctuations.
What is really required for mesoscopic features to emerge in $I(B)$ is that the transport is in the regime of ``slow-hopping'', namely $b_{\s 0} \gg \zeta$. It is in this regime when
sparse PPs, for which  the ${\bm b}_{\s h}$ projections
on the external field  almost coincide, play a distinguished role.

Mesoscopic effect persists when recombination from triplet is also allowed. Important is that the
recombination times from singlet and triplet PPs  differ.

\item
In the  consideration we assumed that the time, $\tau_{\s D}$ of the pair formation is equal to
the time of pair disassembly. This requirement is not restrictive for mesoscopics.
What is important for mesosopics is that both times exceed the recombination time, $\tau$.
In fact, this requirement is a general requirement for spin-dependent recombination, which is at the core of the OMAR effect.

%In the opposite limit, namely $b_{\s 0}\ll \zeta$,
%one can see from  Eqs. (\ref{current}) and (\ref{Delta}) that  OMAR response  in the %strong-field domain has the form $\frac{I(B)-I(\infty)}{I(\infty)}\sim \frac{b_{\s %0}^2}{B^2}$. However Eqs. (\ref{Kparallel}) and (\ref{Kperp}) do not yield mesoscopic %fluctuations in this case. This
%could be expected on general grounds, since $b_{\s 0} \ll \zeta$ means
%that separation of the Zeeman levels is much smaller than their broadening.

\item
As we mentioned above the transport model adopted in the present paper is quite similar
to bipolaron model of transport put forward in  Ref. \onlinecite{Bobbert1}.
Replacement of bipolaron formation by recombination does not bring in any new qualitative features. Thus the  mesoscopic  fluctuations demonstrated in the present paper
can be viewed as a correction due to the local environment to the average current
emerging from the mechanism  Ref.~\onlinecite{Bobbert1}.

\item

Regarding experimental verification of the predicted mesoscopic
fluctuations, we note that there might be an alternative (to decreasing the size) way to
bring samples into a mesoscopic regime. It was demonstrated in Ref. \onlinecite{pillars} that
tin-doped indium oxide (ITO) electrodes exhibit sharp pillars
with areal density of $\sim 1\, \mu\text{m}^{-2}$.
These pillars may cause additional inhomogeneity of the local conductivity
and even define high-conductivity channels in the active layer.
If this is the case, one can estimate from the data in Ref. \onlinecite{pillars},
that a small OLED with area of $\sim 10^{-2}\,\text{cm}^2$ will show mesoscopic fluctuations of $\frac{\delta I}{I} \sim 10^{-3}$.

%As a final remark, note that the slow beating is characterized by
%the time scale $\tau$, the time required for recombination.
%However, {\em transport} characteristics contain $1/\zeta$
%given by Eq. \ref{zeta}, which contains both $\tau$ and $\tau_{\s D}$.
%The latter being the measure of the PP coupling to the environment
%(electrodes). For $\tau_{\s D} \gg \tau$,  $\zeta$ is the
%geometrical mean, $\zeta \sim (\tau \tau_{\s D})^{-1/2}$.

\end{itemize}

\section{Acknowledgements}
We are grateful to E. Erenfreund for useful discussions.
This work was supported by NSF through MRSEC DMR-1121252 and DMR-1104495.

\end{document}